%% file: paper.tex
\documentclass{article}
\usepackage{spconf,amsmath,graphicx,hyperref}

\usepackage{acronym}
\usepackage{orcidlink}
\usepackage{adjustbox}

\usepackage{tikz,pgfplots}
\usepackage{subcaption}

\definecolor{plt_blue_1}{RGB}{30, 119, 180}
\definecolor{plt_orange_1}{RGB}{255, 127, 14}
\definecolor{plt_green_1}{RGB}{44, 160, 44}
\definecolor{plt_red_1}{RGB}{214, 39, 40}
\definecolor{plt_purple_1}{RGB}{148, 103, 189}
\definecolor{plt_pink_1}{RGB}{255, 105, 180}

\usepgfplotslibrary{groupplots}
\pgfplotsset{compat=1.18}

\usetikzlibrary{shapes.geometric, arrows, calc, fit, shapes, backgrounds}

\acrodef{amc}[AMC]{automatic modulation classification}
\acrodef{wgn}[AWGN]{white Gaussian noise}
\acrodef{cfo}[CFO]{center frequency offset}
\acrodef{csp}[CSP]{cyclostationary signal processing}
\acrodef{fsm}[FSM]{frequency-smoothing method}
\acrodef{rf}[RF]{radio frequency}
\acrodef{fc}[FC]{fully connected}
\acrodef{gap}[GAP]{global average pooling}
\acrodef{pca}[PCA]{principal component analysis}
\acrodef{snr}[SNR]{signal-to-noise ratio}
\acrodef{scf}[SCF]{spectral correlation function}
\acrodef{tpr}[TPR]{true positive rate}
\acrodef{fpr}[FPR]{false positive rate}
\acrodef{fp}[FP]{false positive}
\acrodef{roc}[ROC]{receiver operating characteristic}

\title{Approaching domain generalization with embeddings for robust discrimination and recognition of RF communication signals} 
\name{Lukas Henneke~\orcidlink{0009-0006-1391-7178}, Frank Kurth~\orcidlink{0000-0002-9992-083X}}
\address{   Fraunhofer FKIE -- 
			Communication Systems\\
            \{lukas.henneke, frank.kurth\}@fkie.fraunhofer.de}
\begin{document}
%
\maketitle
\begin{abstract}
\Ac{rf} signal recognition plays a critical role in modern wireless communication and security applications. 
Deep learning-based approaches have achieved strong performance but typically rely heavily on extensive training data and often fail to generalize to unseen signals. 
In this paper, we propose a method to learn discriminative embeddings without relying on real-world \ac{rf} signal recordings by training on signals of synthetic wireless protocols. 
We validate the approach on a dataset of real \ac{rf} signals and show that the learned embeddings capture features enabling accurate discrimination of previously unseen real-world signals, highlighting its potential for robust \ac{rf} signal classification and anomaly detection.
 
\end{abstract}
\acresetall

\section{Introduction}
\label{sec:intro}
With the rapid growth of wireless technologies, robust methods for discriminating and recognizing \ac{rf} signals are essential for efficient spectrum utilization, including dynamic spectrum access and cognitive radio, as well as for spectrum monitoring and security applications.
Traditional approaches to \ac{rf} signal classification rely on general signal characteristics, expert-designed features \cite{1998Nandi,2011Clancy}, and \ac{csp} \cite{2000Spooner}, with many methods focusing on the closely related task of \ac{amc}.
In recent years, deep learning has emerged as a powerful paradigm for RF signal classification. 
Early work demonstrates that convolutional neural networks trained directly on raw I/Q data can outperform traditional features in \ac{amc} \cite{2018O’Shea}. 
However, later studies highlight a key limitation: models trained on one dataset typically do not generalize well to datasets collected under different acquisition conditions \cite{2022Latshaw,2023Sathyanarayanan}. 
To address domain shift, hybrid approaches combine deep learning with \ac{csp}, leveraging both data-driven classification and the robustness of traditional features \cite{2022Snoap}. 
Another line of work aims to reduce the risk of domain shift by training on large-scale datasets that cover a broad variety of relevant signal types \cite{2022Boegner,2025Scholl}. 
To further enhance robustness to unseen signals, autoencoder-based novelty detection has been proposed \cite{2023Basak}. 
Although these approaches improve generalization and robustness, they require costly data collection and retraining whenever new signal types emerge. 

These limitations motivate the search for more general and robust signal representations.
In \cite{2017O’Shea}, it is demonstrated that the learned features of intermediate layers of modulation classifiers, i.e., embeddings, form clusters and enable identifying unseen modulations. 
Building on this idea, \cite{2022Persons} propose to use the distributions of learned features to classify new and unseen modulations. 
In \cite{2021Liu}, self-contrastive learning is applied to generate low-dimensional signal representations and enable semi-supervised \ac{amc}. 
Another work combines \ac{csp} with deep learning and metric learning to blindly determine signal modality based on 100-dimensional embeddings \cite{2024Rajagopal}. 
However, these works only focus on modulation or modality classification and are tested only on synthetic data from the same distribution as the training set, which reduces their relevance to the challenges of real-world signal classification.

In this work, we focus on the generation of robust embeddings for discriminating \ac{rf} communication signals.
The goal is to learn a general mapping of signals into a high-dimensional embedding space, such that embeddings of similar signals, i.e., instances of the same wireless protocol, are close, while embeddings of different signal types are well separated.
The embeddings are designed to be robust against realistic impairments, including low signal-to-noise ratio, phase and carrier frequency offsets, and distortions introduced by the transmission channel.
To enable domain generalization across diverse signal types, we rely exclusively on generated signals of synthetic wireless protocols, as collecting datasets for hundreds of real-world wireless protocols is impractical.

Our key contributions are
(1) a methodology to learn robust \ac{rf} signal embeddings that capture similarity across signals of the same wireless protocol while separating different protocols, 
(2) a training strategy based exclusively on synthetic data that avoids the need for large-scale real-world datasets while enabling generalization to unseen protocols,
(3) the curation of an evaluation dataset containing recorded \ac{rf} signals of several wireless protocols, and
(4) a demonstration of the effectiveness of the learned embeddings on the unseen evaluation dataset, showing improved performance over baseline methods and enhanced domain generalization.



\section{RF signal data}
\label{sec:data}

\subsection{Synthetic training data}
The waveform characteristics of wireless protocols are primarily determined by the design of the physical layer. 
In addition to parameters such as modulation type, symbol rate, and pulse shaping, recurring bit patterns inserted for synchronization, equalization, or frame signaling yield distinctive signal features that can be exploited to recognize the underlying protocol.
In this work, we therefore aim to simulate wireless protocols at the level of the physical layer together with their frame structures. 
For this purpose, we consider single-carrier and multi-carrier modulation schemes as well as spread spectrum techniques. 
ASK, PSK, APSK, QAM and M-FSK with varying modulation orders are used for single-carrier modulation, whereas OFDM, with PSK and QAM to modulate the carriers, and FDM using PSK, QAM and FSK2 are used for multi-carrier modulation. 
Additionally, we consider the spread spectrum techniques CSS and DSSS, the latter employing Barker and PN~sequences with PSK and QAM for modulation.
A random protocol instance is generated by first selecting the modulation scheme and its associated parameters. 
Next, it is determined whether successive frames are transmitted in bursts or continuously.
The frame structure consists of a synchronization sequence, frame header, and payload, each of which has a fixed length in continuous mode, whereas the length of the payload may vary in burst mode.
Furthermore, additional synchronization sequences may be inserted within or at the end of the frame in burst mode.
For synchronization, all-zero sequences, all-one sequences, alternating zero-one sequences, or repetitions of longer bit patterns are used.
In the special case of OFDM, only a subset of carriers may be active during synchronization, or alternatively, Zadoff-Chu sequences are employed.
The frame header is characterized by a subset of bits that are randomly set and remain fixed across all generated frames of the protocol.
The payload is generated entirely at random. 
The symbol rate is chosen such that the resulting bandwidth is a fraction of the target sample rate to which the signal is resampled after its generation.
For generating a signal instance for training, multiple frames are concatenated (including pauses in burst mode) until the required number of samples is obtained.
Finally, to simulate transmission impairments, we use a random phase offset, introduce a normally distributed \ac{cfo} with standard deviation $\sigma_\mathrm{RFO}$ relative to the signal sampling rate, apply one of the 3GPP TDL channel models A–E \cite{3gpp38901}, and add \ac{wgn}.
For demonstration purposes, we upload a dataset comprising signal instances of numerous simulated protocols.\footnotemark[1]

\subsection{Real-world evaluation data}
In order to evaluate the proposed approach in a realistic context, we curate a dataset that contains measured I/Q data of various wireless protocols.
The evaluation dataset is constructed by extracting single bursts from wideband recordings from three publicly available data sources \cite{2020Vuorenmaa, 2024Basak, iqengine}.
The resulting collection of signals covers several \ac{rf} communication protocols, most of them operating in ISM bands, and predominantly includes drone remote control (RC) and video transmission link (VL) signals, in addition to 
protocols such as WiFi and Bluetooth.
For each signal type, a fixed target sample rate is chosen that slightly exceeds the occupied bandwidth and is used for downconversion. 
All 18 signal types, together with their corresponding sample rates $f_s$, the number of different emitters $N_e$ in the dataset, and the originating data sources are summarized in Table~\ref{tab:signals}.
Signal types are named according to the protocol name whenever the underlying communication standard could be identified.
Otherwise, the corresponding drone model serves as the label.
Note that the WiFi label refers to all OFDM-based WiFi signals (802.11a/g/n/…) with a bandwidth of 20 MHz, whereas WiFi-DSSS denotes signal instances of the 802.11b standard.
We extract 50 bursts per signal type, except for RFD900p where only 30 bursts were available, yielding a total dataset size of 880 signal instances.
The number of samples per signal instance varies both within and across classes, ranging from a minimum of 216 samples observed for a Bluetooth signal to a maximum of 123,228 samples for a Lightbridge VL signal.
Before signal extraction, noise floor and \ac{snr} are estimated, such that only bursts with an in-band \ac{snr} greater than 23~dB are included. 
For evaluation, this enables variation of the \ac{snr} by adding \ac{wgn}. 
Further data augmentations include applying a random phase and carrier frequency offset as well as simulating channel distortions using the 3GPP TDL channel models \cite{3gpp38901}.
The constructed dataset is made publicly available to facilitate reproducibility and enable comparative evaluations.\footnote{\scriptsize \url{https://fordatis.fraunhofer.de/handle/fordatis/460}}

\begin{table}
\centering
\caption{Summary of the evaluation dataset.}
\label{tab:signals}
\scalebox{1.0}{
\tiny
\setlength{\tabcolsep}{2.5pt}
\begin{tabular*}{0.88\columnwidth}{l l l c l | l l l c l}

\hline
 & Signal type & $f_s$ & $N_{e}$ & Source & & Signal type & $f_s$ & $N_{e}$ & Source \\
\hline
1  & Bluetooth        & 2 MHz   & 1 & \cite{iqengine} & 10 & OcuSync VL       & 20 MHz  & 4 & \cite{2020Vuorenmaa,2024Basak}\\
2  & Bluetooth LE     & 2 MHz   & 3 & \cite{2020Vuorenmaa,2024Basak} & 11 & Q205 RC          & 2 MHz   & 1 & \cite{2024Basak}\\
3  & DECT6            & 2 MHz   & 1 & \cite{iqengine} & 12 & RFD900p          & 0.5 MHz & 1 & \cite{iqengine}\\
4  & DroneID          & 12 MHz  & 4 & \cite{2020Vuorenmaa,2024Basak} & 13 & SJRC RC          & 2 MHz   & 1 & \cite{2024Basak}\\
5  & FrSky            & 0.5 MHz & 1 & \cite{2024Basak} & 14 & Spektrum DSMX    & 2 MHz   & 1 & \cite{2024Basak}\\
6  & Lightbridge RC   & 2 MHz   & 6 & \cite{2020Vuorenmaa,2024Basak} & 15 & WLtoys RC        & 2 MHz   & 1 & \cite{2024Basak}\\
7  & Lightbridge VL   & 12 MHz  & 6 & \cite{2020Vuorenmaa,2024Basak} & 16 & WiFi             & 20 MHz  & 7 & \cite{2020Vuorenmaa,2024Basak}\\
8  & NineEagles RC    & 2 MHz   & 1 & \cite{2024Basak} & 17 & WiFi-DSSS        & 20 MHz  & 2 & \cite{2020Vuorenmaa}\\
9  & OcuSync RC       & 2 MHz   & 4 & \cite{2020Vuorenmaa,2024Basak} & 18 & Yuneec RC        & 5 MHz   & 1 & \cite{2020Vuorenmaa}\\
\hline
\end{tabular*}
}

\end{table}

\section{Methodology}
\label{sec:method}

The methods considered in this work are based on features 
solely extracted from I/Q data, i.e., no additional information such as center frequency, sample rate or bandwidth is used to differentiate signals. 
Therefore, it is important that signals of the same type are provided at the same sample rate, as the corresponding features may differ for different sample rates. 

\subsection{Statistical modulation features}
The first baseline method utilizes signal features that were originally proposed for \ac{amc}. 
We employ the feature set listed in Table 2 of \cite{2018O’Shea}, which includes the 9 Nandi–Azzouz features \cite{1998Nandi}, 10 higher-order moments, and 7 additional higher-order cumulants, resulting in a 26-dimensional feature vector.
Note that several of these features span different value ranges. 
However, normalization was found to degrade discriminative performance.
Consequently, the features are used in their raw form, 
while acknowledging that suitable feature scaling may improve separability in feature space.

\subsection{CSP-based feature extraction}
Since the signals used for evaluation are digitally modulated and exhibit cyclostationary properties, we consider \ac{csp}-based features as a second baseline.
We use the magnitude of the \ac{scf}, which has proven to be a promising candidate as it reveals distinctive characteristics across the considered signal types.
The \ac{scf} is estimated using the \ac{fsm} \cite{1994Spooner} with coarse resolution, 
based on an FFT size of $2^{14}$ and a frequency-smoothing window of length 256, which yields 64 bins in the frequency domain.
The cyclic frequency is evaluated over the range $[0, 0.5]$ with step size $0.01$, resulting in 50 bins in the cyclic frequency domain.
Consequently, each signal instance is represented by a $64 \times 50$-dimensional \ac{scf} matrix.
To further reduce dimensionality, a \ac{pca} transformation is computed from the collection of all \ac{scf} matrices and subsequently applied to each instance, resulting in 128-dimensional feature vectors.

\subsection{Learning robust \ac{rf} signal embeddings}

\begin{figure}[t]
\centering
\begin{adjustbox}{max width=0.5\textwidth,right=8.0cm}
\input{img/model.tex}
\end{adjustbox}
\vspace{-0.2cm}
\caption{Network design for transforming the signal representation $\text{x}$ into an embedding $\text{z}$ and classification probabilities $\text{y}$. Dimensions are shown for the case of a 2D input.}
\vspace{-0.3cm}
\label{fig:network}
\end{figure}
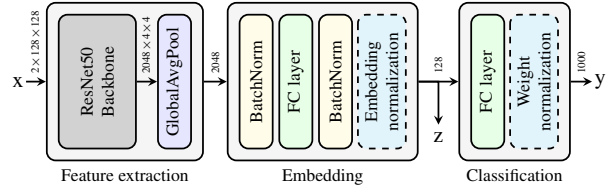

Inspired by recent deep learning approaches in face recognition \cite{2022Deng}, we formulate a classification task to learn embeddings that capture discriminative features of \ac{rf} signals. 
We consider two types of input representations: raw I/Q samples as 1D input and a time–frequency representation as 2D input. 
For training, we generate signals of length $2^{14}$ samples, resulting in an input vector of size $2\times16384$ for the 1D case. 
For the 2D case, we apply an STFT with non-overlapping segments, a rectangular window, and an FFT size of 128, yielding an input of size $2\times128\times128$. 
In both cases, the two channels represent the real and imaginary parts.
As feature extractor, we employ ResNet50 \cite{2016He}. 
Between the \ac{gap} layer and the final \ac{fc} layer for classification, we insert additional layers producing the embedding (see Fig.~\ref{fig:network}). 
Since ResNet50 is originally \mbox{designed} for 2D inputs, we adapt it to the 1D case by replacing all 2D layers with their 1D counterparts, resulting in models with 
16.4~M (1D) and 23.9~M (2D) parameters. 
For training, we adopt ArcFace (AF), an additive angular margin loss known to enhance the discriminative power of learned embeddings, and compare it against Norm-Softmax (NS) and Softmax (Sm) \cite{2022Deng}. 
Note that embedding and weight normalization shown in Fig.~\ref{fig:network} are applied only for Norm-Softmax and ArcFace, which distribute embeddings on a hypersphere of radius $s$. 
In these cases, no bias is used in both \ac{fc} layers.
Preliminary experiments indicated that setting $s=8$ yields stable training. 
Furthermore, an embedding size of 128 and an angular margin penalty of $m=0.5$ proved to be reasonable choices.
We employ 1000 synthetic \ac{rf} protocols for training. 
Both 1D and 2D versions of the modified ResNet50, abbreviated as RN-1D and RN-2D, are trained to classify these protocols with the three loss functions, resulting in six model variants in total. 
We use SGD optimizer with momentum 0.9 and weight decay $5\mathrm{e}{-4}$. 
The learning rate is initialized to 0.025 and reduced by a factor of 10 after epochs 15 and 18. 
Training is performed for 20 epochs with a batch size of 128, where 50 signal instances per protocol are generated per epoch, with \ac{snr} between 3 and 30~dB and $\sigma_\mathrm{RFO}=0.02$. 
We perform ten training runs and report mean and maximum performance. 
Each run uses a newly generated set of 1000 RF protocols, applied consistently across all six model variants. 
For inference, the fully convolutional ResNet backbone with the subsequent \ac{gap} layer allows to process variable-length signals, such that a single embedding is produced per evaluation signal.

\section{Evaluation}
\begin{figure*}[!ht]
    \centering
    \begin{subfigure}[b]{0.3\textwidth}
        \caption{Statistical modulation features}
        \includegraphics[width=\linewidth]{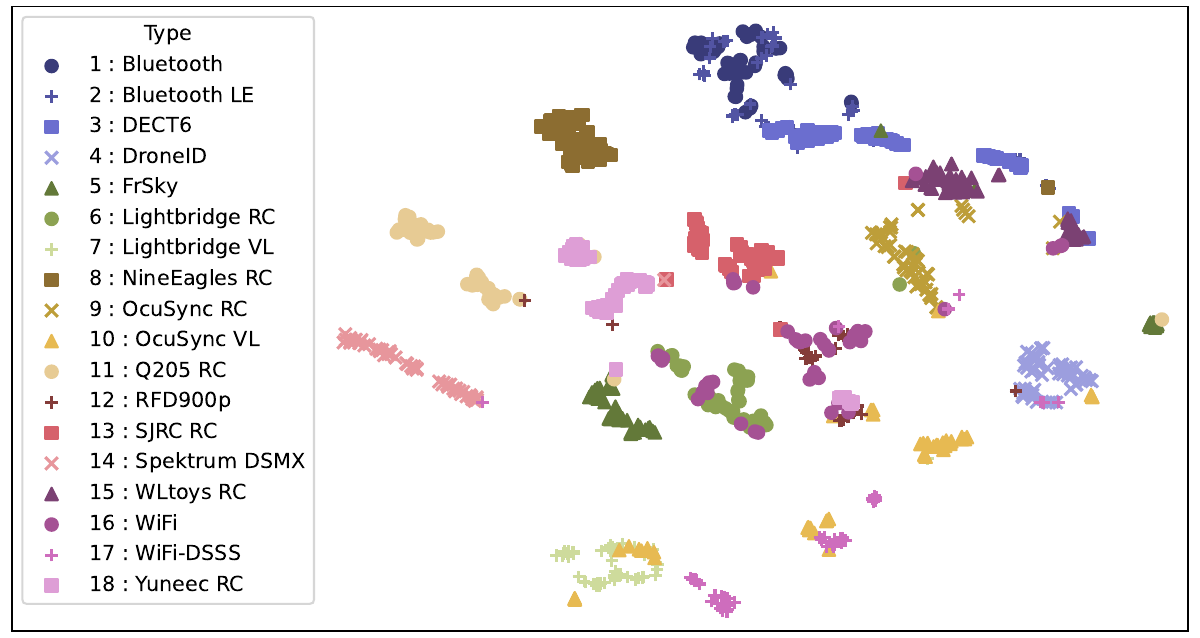}
    \end{subfigure}
    \hfill
    \begin{subfigure}[b]{0.3\textwidth}
        \caption{SCF-PCA features}
        \includegraphics[width=\linewidth]{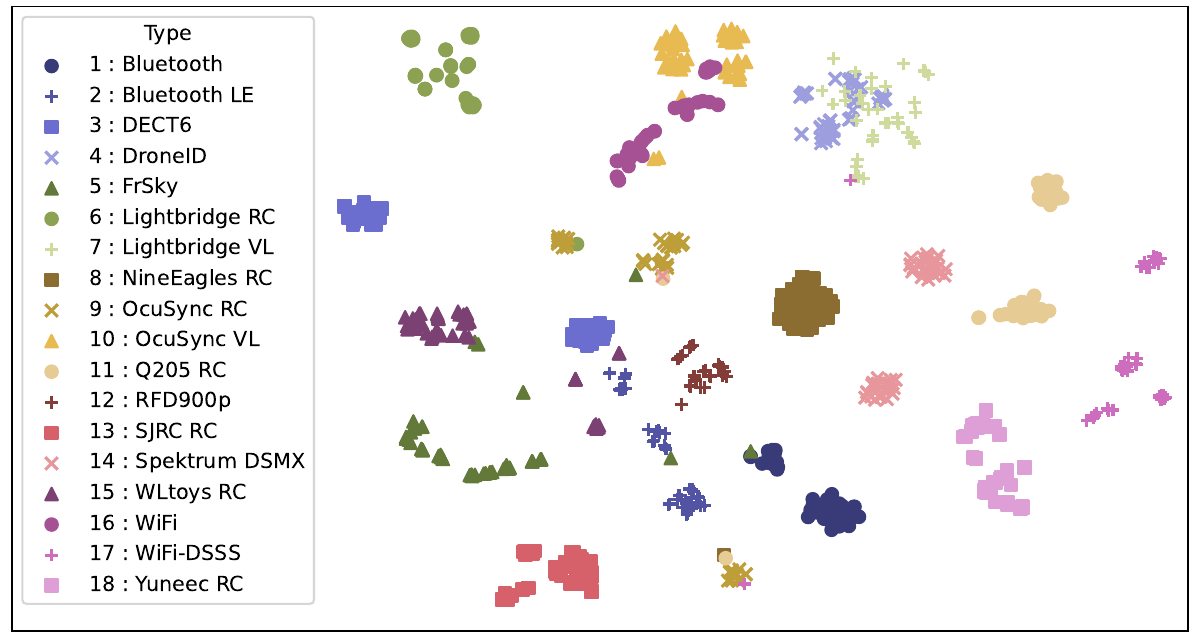}
    \end{subfigure}
    \hfill
    \begin{subfigure}[b]{0.3\textwidth}
        \caption{Learned embeddings}
        \includegraphics[width=\linewidth]{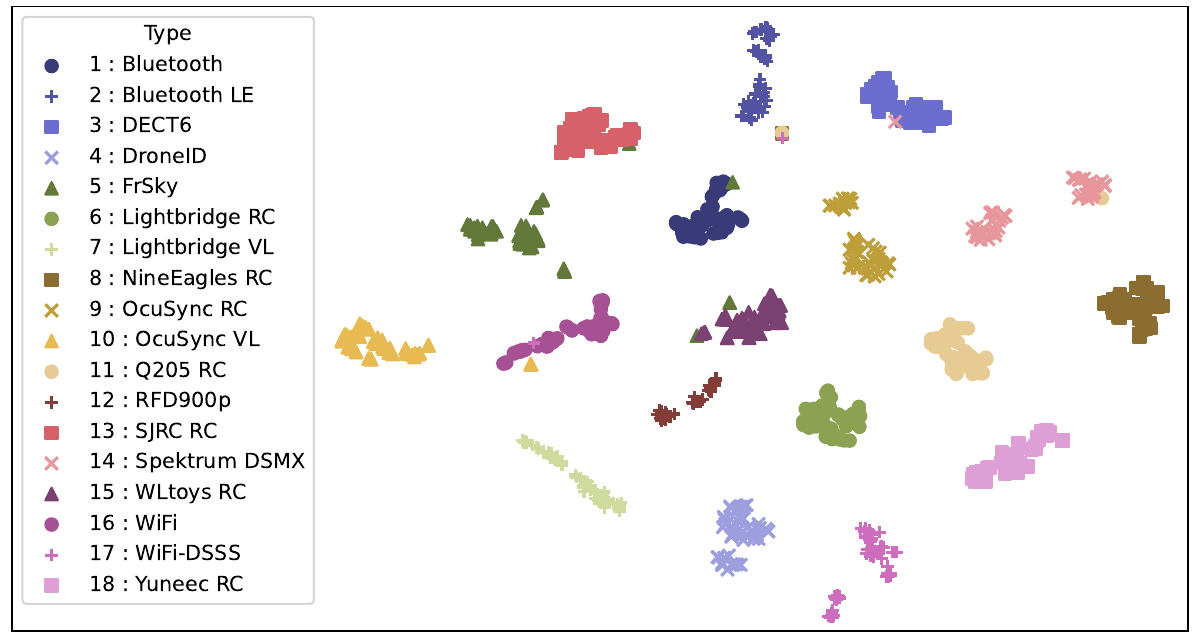}
    \end{subfigure}
    \vspace{-0.05cm}
    \caption{Two-dimensional t-SNE projection of feature representations of the evaluation dataset at in-band SNR = 10 dB.}
    \vspace{-0.2cm}
    \label{fig:tsne}
\end{figure*}
\label{sec:eval}

\begin{figure}[!ht]
    \centering
    	\centerline{
    	\resizebox{\columnwidth}{!}{\input{img/plots}}
    	}
        \vspace{-0.3cm}
    	\caption{Robustness against \ac{wgn} and \ac{cfo} impairments.}
	   \label{fig:plots}
\end{figure}
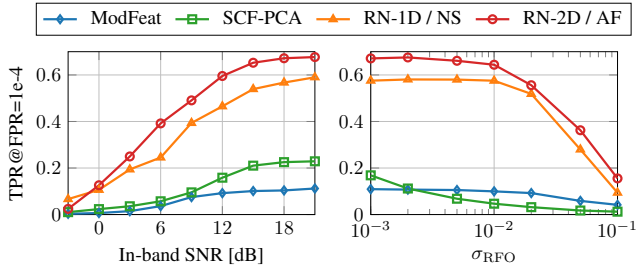

To assess the intra-class compactness and inter-class separability induced by the feature vectors and embeddings, we consider pairwise distances in the feature space, i.e., we perform one-to-one signal verification.
Each pair of signals is treated as a binary classification problem, depending on whether the two signals belong to the same class or to different classes.
Pairs with a distance below a threshold are considered positive.
Performance is quantified using the \ac{tpr} and \ac{fpr} as functions of the distance threshold. 
Since the number of different-class pairs (365,500) exceeds the same-class pairs (21,260) by far, the binary classification task is highly imbalanced. 
As even a few \acp{fp} can be critical in practical applications, we report \ac{tpr} at fixed \ac{fpr} values of $1\text{e}{-3}$ and $1\text{e}{-4}$.
For example, at a \ac{tpr} of 0.5 and an \ac{fpr} of $1\mathrm{e}{-4}$, 10,630 positive pairs are correctly identified with just 39 \acp{fp}, which already allows effective clustering with minimal errors.

\subsection{Feature and embedding analysis}

Fig.~\ref{fig:tsne} visualizes the distributions of the evaluation dataset in the feature space using the baseline methods and an embedding variant. 
All three methods form several distinct clusters, confirming their applicability to the task, while learned embeddings achieve the most pronounced separation. 
Table~\ref{tab:results} reports the \acp{tpr} of baseline methods and embedding variants on the non-impaired dataset. 
Embeddings substantially surpass the baselines, with 2D-based embeddings consistently outperforming 1D-based ones. 
Performance is further shaped by the loss function.
For 1D input, Norm-Softmax yields the best results, whereas ArcFace surprisingly performs worst. 
For 2D input, ArcFace achieves the most effective embeddings. 
Fig.~\ref{fig:plots} shows that the embeddings maintain \mbox{acceptable} \acp{tpr} at decreasing \ac{snr} and increasing \ac{cfo} variability, demonstrating their robustness. 
In contrast, the strict \ac{fpr} requirements lead to poor performance of the baseline features.

\begin{table}[h]
\centering
\caption{Performance of baseline and embedding methods.}
\label{tab:results}
\vspace{-0.1cm}
\scalebox{1.0}{
\scriptsize
\begin{tabular}{l|cc|cc}
 & \multicolumn{2}{c}{TPR@FPR=$10^{-3}$} & \multicolumn{2}{c}{TPR@FPR=$10^{-4}$} \\
\hline
\textbf{Method} & \textbf{Mean $\pm$ Std} & \textbf{Max} & \textbf{Mean $\pm$ Std} & \textbf{Max} \\
\hline
ModFeat & 0.168 & 0.168 & 0.118 & 0.118 \\
SCF-PCA & 0.258 & 0.258 & 0.238 & 0.238 \\
RN-1D / Sm & 0.463 $\pm$ 0.045 & 0.518 & 0.413 $\pm$ 0.051 & 0.489 \\
RN-1D / NS & 0.606 $\pm$ 0.035 & 0.652 & 0.519 $\pm$ 0.052 & 0.585 \\
RN-1D / AF & 0.387 $\pm$ 0.066 & 0.488 & 0.304 $\pm$ 0.076 & 0.396 \\
RN-2D / Sm & 0.637 $\pm$ 0.075 & 0.732 & 0.566 $\pm$ 0.089 & 0.680 \\
RN-2D / NS & 0.697 $\pm$ 0.047 & 0.778 & 0.573 $\pm$ 0.052 & 0.687 \\
RN-2D / AF & 0.685 $\pm$ 0.055 & \textbf{0.793} & 0.582 $\pm$ 0.053 & \textbf{0.693} \\
\hline
\end{tabular}}
\vspace{-0.2cm}
\end{table}

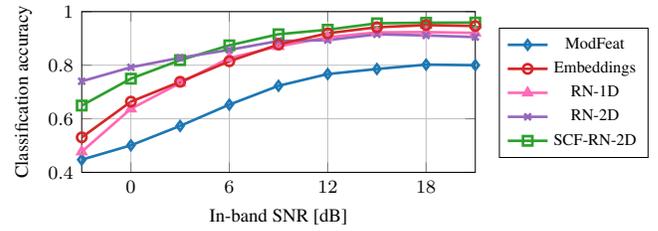
\begin{figure}[!ht]
    \centering
    	\centerline{
    	\resizebox{\columnwidth}{!}{\input{img/clsres}}
    	}
        \vspace{-0.3cm}
    	\caption{Comparison of classification results under \ac{wgn}, \ac{cfo} ($\sigma_\mathrm{RFO} = 0.02$), and simulated wireless channels.}
	   \label{fig:clsres}
\end{figure}

\subsection{Downstream classification task}
To further investigate the practical utility of the baseline features and learned embeddings, we consider a downstream classification task.
The evaluation dataset is split into training (54\%) and test (46\%) sets across its 18 classes, with the previously described augmentations applied to expand both sets.
We train five classifiers in total.
First, a classifier composed of two \ac{fc} layers of size 128 and a final classification layer of size 18 is trained using either the 26-D modulation features or the \mbox{128-D} embeddings of the best-performing model (RN-2D/AF) as input.
In addition, RN-1D and RN-2D are trained from scratch on the corresponding input representations to learn discriminative features directly from real-world training data.
Finally, \ac{scf} matrices without \ac{pca} 
are utilized as 2D input to train another ResNet classifier.
The classification results in Fig.~\ref{fig:clsres} show that the embedding-based classifier performs on par with the other approaches and is only surpassed by the \ac{scf}-based classifier and, at low \ac{snr}, by the 2D ResNet.
This suggests that the embeddings already encode most signal characteristics, leaving limited scope for improvement via direct feature learning on real-world data.



\section{Conclusion}
We present a methodology for learning \ac{rf} signal embeddings exclusively from synthetic training data, avoiding the need for extensive field-collected datasets. 
The embeddings generalize well to real-world signals of practical wireless protocols and remain robust under realistic impairments, achieving superior clustering and competitive classification performance.
Future work includes deeper analysis of the encoded features, integration of \ac{csp}-based features, and exploration of improved architectures and training strategies. 
We will also extend evaluation to more diverse real-world datasets and study embeddings for applications such as open-set \ac{rf} signal recognition.


\bibliographystyle{IEEEbib}
\bibliography{refs}

\end{document}

%% file: img/model.tex
\usetikzlibrary{arrows, fit}
\tikzset{inv/.style={fill=gray!8, rectangle, node distance=1cm, minimum width=2.8cm, text centered, minimum height=1.4em, text width=1.9cm}}
\tikzset{rn/.style={fill=gray!35, draw, rectangle, rounded corners, thick, node distance=1cm, minimum width=1.1cm, text centered, minimum height=3.5em, text width=1.9cm, font=\footnotesize}}
\tikzset{bn/.style={fill=yellow!10, draw, rectangle, rounded corners, thick, node distance=1cm, minimum width=1.1cm, text centered, minimum height=1.5em, text width=1.9cm, font=\footnotesize}}
\tikzset{dns/.style={fill=green!10, draw, rectangle, rounded corners, thick, node distance=1cm, minimum width=1.9cm, text centered, minimum height=1.5em, text width=1.9cm, font=\footnotesize}}
\tikzset{nrm/.style={fill=cyan!10, draw, dashed, rectangle, rounded corners, thick, node distance=1cm, minimum width=1.9cm, text centered, minimum height=1.5em, text width=1.9cm, font=\footnotesize}}
\tikzset{gap/.style={fill=blue!10, draw, rectangle, rounded corners, thick, node distance=1cm, minimum width=1.9cm, text centered, minimum height=1.5em, text width=1.9cm, font=\footnotesize}}
\tikzset{sigm/.style={fill=red!10, draw, rectangle, rounded corners, thick, node distance=1cm, minimum width=2.3cm, text centered, minimum height=1.5em, text width=2.8cm}}
\tikzset{convtr/.style={fill=yellow!10, draw, rectangle, rounded corners, thick, node distance=1cm, minimum width=2.3cm, text centered, minimum height=1.5em, text width=2.8cm}}

\tikzset{container/.style={fill=gray!8, draw, rectangle, rounded corners, thick, inner xsep=0.5em,inner ysep=0.5em, minimum width=0.2cm}}

\tikzset{font={\fontsize{10pt}{10}\selectfont}}
\tikzstyle{arrow} = [thick, ->, >=stealth]

\begin{tikzpicture}
\node(input)[text centered,inner sep=1pt] {$\text{x}$};

\node(resnet)[rn,node distance=3.6em, right of=input, rotate=90]{ResNet50 Backbone};
\node(glob)[gap,node distance=3.5em, right of=resnet, rotate=90]{GlobalAvgPool};
\node(bn1)[bn,node distance=3.7em, right of=glob, rotate=90]{BatchNorm};
\node(emb)[dns,node distance=1.8em, right of=bn1, rotate=90]{FC layer};
\node(bn2)[bn,node distance=1.8em, right of=emb, rotate=90]{BatchNorm};
\node(nrm1)[nrm,node distance=2.1em, right of=bn2, rotate=90]{Embedding normalization};
\node(cls)[dns,node distance=4.8em, right of=nrm1, rotate=90]{FC layer};
\node(nrm2)[nrm,node distance=2.1em, right of=cls, rotate=90]{Weight\\normalization};
\node(output)[text centered,node distance=3.0em, right of=nrm2,inner sep=1pt] {$\text{y}$};

\begin{scope}[on background layer]
\node(feat)[container, fit=(resnet)(glob)] {};
\node at (feat.south) [text centered,below,node distance=0 and 0, align=center,xshift=0cm] (backendtxt) {\footnotesize Feature extraction};
\node(embed)[container, fit=(bn1)(nrm1)] {};
\node at (embed.south) [text centered,below,node distance=0 and 0, align=center,xshift=0cm] (backendtxt) {\footnotesize Embedding};
\node(class)[container, fit=(cls)(nrm2)] {};
\node at (class.south) [text centered,below,node distance=0 and 0, align=center,xshift=0cm] (backendtxt) {\footnotesize Classification};
\end{scope}

\draw[arrow](input) -- node[right, rotate=90] {\tiny 2$\times$128$\times$128} (feat);
\draw[arrow](resnet) -- node[right, rotate=90] {\tiny 2048$\times$4$\times$4} (glob);
\draw[arrow](feat) -- node[right, rotate=90] {\tiny 2048} (embed);
\draw[arrow](embed) -- node[right, rotate=90] {\tiny 128} (class);
\node at (embed.east) [yshift=-2.5em, xshift=0.9em, inner sep=3pt] (z) {$\text{z}$};
\draw[arrow](embed.east) -| (z);
\draw[arrow](class) -- node[right, rotate=90] {\tiny 1000} (output);

\end{tikzpicture}

%% file: img/plots.tex
\begin{tikzpicture}
\begin{groupplot}[
  group style={
    group size=2 by 1,
    horizontal sep=1cm,
  },
  width=6cm,
  height=4.5cm,
]

\nextgroupplot[
  xlabel={In-band SNR [dB]},
  ylabel={TPR@FPR=1e-4},
  xmin=-3, xmax=21,
  ymin=0, ymax=0.7,
  xtick={-10,0,6,12,18},
  ytick={0,0.2,...,1.0},
  grid=major,
]
\addplot+[mark=diamond, color=plt_blue_1, line width=0.4mm] coordinates {(-3, 0.00323142) (0, 0.00651929) (3, 0.01457667) (6, 0.03626999) (9, 0.07507526) (12, 0.09206491) (15, 0.10104892) (18, 0.10367827) (21, 0.11196143)};
\addplot+[mark=square, color=plt_green_1, line width=0.4mm] coordinates {(-3, 0.01035748) (0, 0.02342427) (3, 0.03568674) (6, 0.05708843) (9, 0.09539511) (12, 0.15842427) (15, 0.21020226) (18, 0.22528692) (21, 0.22891816)};
\addplot+[mark=triangle, color=plt_orange_1, line width=0.4mm] coordinates {(-3, 0.0664064) (0, 0.1061524) (3, 0.19396049) (6, 0.2451223) (9, 0.39359831) (12, 0.46514111) (15, 0.53916745) (18, 0.56722954) (21, 0.58992944)};
\addplot+[mark=o, color=plt_red_1, line width=0.4mm] coordinates {(-3, 0.02429445) (0, 0.12594073) (3, 0.24970367) (6, 0.39175917) (9, 0.49073848) (12, 0.59547037) (15, 0.65246943) (18, 0.67181091) (21, 0.67690028)};

\nextgroupplot[
  xlabel={$\sigma_\mathrm{RFO}$},
  xmin=0.001, xmax=0.1,
  ymin=0, ymax=0.7,
  xmode=log,
  xtick={0,0.001,0.01,0.1},
  ytick={0,0.2,...,1.0},
  grid=major,
  legend style={at={(-0.15,1.3)}, cells={align=center}, column sep=1ex, anchor=north, legend columns=4},
]
\addplot+[mark=diamond, color=plt_blue_1, line width=0.4mm] coordinates {(0.0, 0.11854186) (0.001, 0.10916745) (0.002, 0.10724835) (0.005, 0.10537629) (0.01, 0.09966134) (0.02, 0.09232832) (0.05, 0.05871119) (0.1, 0.04168391)};
\addplot+[mark=square, color=plt_green_1, line width=0.4mm] coordinates {(0.0, 0.23976482) (0.001, 0.16860301) (0.002, 0.11194732) (0.005, 0.06800564) (0.01, 0.04662747) (0.02, 0.03182502) (0.05, 0.01737065) (0.1, 0.01237065)};
\addplot+[mark=triangle, color=plt_orange_1, line width=0.4mm] coordinates {(0.0, 0.60231891) (0.001, 0.57542803) (0.002, 0.5806397) (0.005, 0.58005644) (0.01, 0.57523048) (0.02, 0.51851834) (0.05, 0.27956726) (0.1, 0.09389934)};
\addplot+[mark=o, color=plt_red_1, line width=0.4mm] coordinates {(0.0, 0.68977422) (0.001, 0.67112888) (0.002, 0.675254) (0.005, 0.66091251) (0.01, 0.64390875) (0.02, 0.55594073) (0.05, 0.36232361) (0.1, 0.15534337)};

\legend{ModFeat, SCF-PCA, RN-1D / NS, RN-2D / AF}


\end{groupplot}
\end{tikzpicture}

%% file: img/clsres.tex
\begin{tikzpicture}
\begin{axis}[
    width=7.5cm, height=4cm, 
    xlabel={In-band SNR [dB]}, ylabel={Classification accuracy},
    grid=major,
  xmin=-3, xmax=21,
  ymin=0.4, ymax=1.0,
  xtick={-10,0,6,12,18},
  ytick={0,0.2,...,1.0},
  label style={font=\footnotesize}, 
    tick label style={font=\footnotesize}, 
    legend style={font=\scriptsize, at={(1.25,0.9)}, anchor=north, legend columns=1, cells={align=left}} 
]

\addplot+[mark=diamond, color=plt_blue_1, line width=0.4mm, forget plot] coordinates {
    (-3, 0.44722222) (0, 0.50083333) (3, 0.57333333) (6, 0.65277778) (9, 0.72305556) (12, 0.76694444) (15, 0.78583333) (18, 0.80222222) (21, 0.8)
};

\addplot+[mark=square, color=plt_green_1, line width=0.4mm, forget plot] coordinates {
    (-3, 0.64944444) (0, 0.74972222) (3, 0.81888889) (6, 0.87416667) (9, 0.91555556) (12, 0.9325) (15, 0.95638889) (18, 0.95861111) (21, 0.95916667)
};

\addplot+[mark=triangle, color=plt_pink_1, line width=0.4mm, forget plot] coordinates {
    (-3, 0.47694444) (0, 0.63722222) (3, 0.73333333) (6, 0.82666667) (9, 0.87027778) (12, 0.90416667) (15, 0.9225) (18, 0.92361111) (21, 0.92055556)
};

\addplot+[mark=x, color=plt_purple_1, line width=0.4mm, forget plot] coordinates {
    (-3, 0.74) (0, 0.79222222) (3, 0.82777778) (6, 0.85722222) (9, 0.89055556) (12, 0.89361111) (15, 0.91527778) (18, 0.91111111) (21, 0.905)
};

\addplot+[mark=o, color=plt_red_1, line width=0.4mm, forget plot] coordinates {
    (-3, 0.53055556) (0, 0.66388889) (3, 0.73805556) (6, 0.81472222) (9, 0.87722222) (12, 0.91916667) (15, 0.94166667) (18, 0.94972222) (21, 0.94638889)
};

\addlegendimage{mark=diamond, color=plt_blue_1, line width=0.4mm}
\addlegendentry{ModFeat}

\addlegendimage{mark=o, color=plt_red_1, line width=0.4mm}
\addlegendentry{Embeddings}

\addlegendimage{mark=triangle, color=plt_pink_1, line width=0.4mm}
\addlegendentry{RN-1D}

\addlegendimage{mark=x, color=plt_purple_1, line width=0.4mm}
\addlegendentry{RN-2D}

\addlegendimage{mark=square, color=plt_green_1, line width=0.4mm}
\addlegendentry{SCF-RN-2D}

\end{axis}
\end{tikzpicture}